\documentclass[manuscript]{aastex631}
\usepackage{latexsym}

\shorttitle{Magma Oceans on Lava Planets}
\shortauthors{Boukar\'e et al.}
\graphicspath{{./}{figures/}}
\begin{document}

\title{Deep two-phase, hemispherical magma oceans on lava planets}

\author[0000-0002-9265-4209]{Charles-\'Edouard Boukar\'e}
\affiliation{Institut de Physique du Globe de Paris,
1, rue Jussieu,
Paris CEDEX 05, 75238, France}

\author[0000-0001-6129-5699]{Nicolas B. Cowan}
\affiliation{Department of Earth and Planetary Sciences, McGill University,
3450, rue University,
 Montréal, Québec, H3A 0E8, Canada}
\affiliation{Department of Physics, McGill University,
3600, rue University,
 Montréal, Québec, H3A 2T8, Canada}
 
\author[0000-0001-9337-4789]{James Badro}
\affiliation{Institut de Physique du Globe de Paris,
1, rue Jussieu,
Paris CEDEX 05, 75238, France}



\begin{abstract}

Astronomers have discovered a handful of exoplanets with rocky bulk compositions but orbiting so close to their host star that the surface of the planet must be at least partially molten.  It is expected that the dayside of such ``lava planets'' harbors a rock vapor atmosphere that flows quickly towards the airless nightside---this partial atmosphere is critical to the interpretation of lava planet observations, but transports negligible heat towards the nightside. As a result, the surface temperature of the magma ocean may range from 3000~K near the sub-stellar point down to 1500~K near the day--night terminator.  We use simple models incorporating the thermodynamics and geochemistry of partial melt to predict the physical and chemical properties of the magma ocean as a function of the distance from the sub-stellar point. Our two principal findings are that 1) the dayside magma ocean is much deeper than previously thought, probably extending down to the core--mantle boundary in some locations, and 2) much of the dayside is only partially molten, leading to gradients in the surface chemistry of the magma ocean.  These findings have important implications for the dynamics of the magma ocean as well as the composition and dynamics of the atmosphere. 

\end{abstract}



  \section{Introduction} \label{sec:intro}

Astronomers have recently discovered a new class of planet, planets with bulk densities suggestive of terrestrial composition, but orbiting so close to their host star that the dayside equilibrium temperature exceeds the solidus temperature of silicates, and hence must be molten.  Notable examples include CoRoT-7b \citep{leger2009}, Kepler-10b \citep{2011ApJ...729...27B}, 55~Cnc~e \citep{Demory2011,Winn2011}, and K2-141b \citep{2018A&A...612A..95B,2018AJ....155..107M}. 
%\footnote{While 55~Cancri e is a relatively small ultra hot planet \citep{}, the latest estimates of its mass and radius suggest that it probably harbours a significant gas envelope \citep{Bourrier2018Morris2021} and hence is not a lava planet.}  
These ``lava planets'' are predicted to have a permanent magma ocean with an overlying rock-vapor atmosphere \citep{2009ApJ...703L.113S}.  As with most short-period exoplanets, lava planets are expected to be tidally-locked into synchronous rotation \citep{leger2011}.  For a review of lava planets, read \cite{2021ChEG...81l5735C}.  

\subsection{Atmospheres of Lava Planets}
In the likely event that a lava planet has already lost its volatiles to space, then it cannot maintain a steady-state atmosphere on its nightside \citep{leger2011}. Instead, it will have a partial atmosphere of rock vapor on the dayside that blows towards the airless nightside, cools and condenses onto the surface, only to be returned to the sub-stellar point via magma ocean currents and solid-state flow near the shores of the ocean \citep{castan2011,kite2016,nguyen2020}.  If a lava planet somehow retains volatiles over gigayears, then it may have a global atmosphere and relatively homogenized surface temperatures \citep{2017ApJ...849..152H}.  In the current paper we adopt the majority view, namely that lava planets have a day-to-night temperature contrast of more than 2000~K, with a dayside magma ocean and an airless nightside.    

Fortunately, the high temperatures of lava planets make them amenable to observational studies.  Transits, eclipses and phase variations of lava planets have been observed with the CoRoT, Kepler, and Spitzer space telescopes \citep{leger2009,batalha2011,2016Natur.532..207D}, and further observations have been approved for the Hubble and James Webb space telescopes \citep[e.g.,][]{2021jwst.prop.2347D,2021jwst.prop.2159E,2021jwst.prop.1952H,2021hst..prop16660Q}. These measurements constrain the surface temperature of a lava planet as a function of longitude, the atmospheric composition, and even the atmospheric temperature structure \citep{2020MNRAS.494.1490Z,2021MNRAS.500.2197Z,nguyen2020}.  While remote sensing can only directly probe an exoplanet's atmosphere and surface, one may sometimes indirectly infer properties of its interior \citep{2014EOSTr..95..209C}.

The atmospheric composition and dynamics on a lava planet depend on the composition and dynamics of its magma ocean.  To first approximation, the atmosphere at some location on a lava planet is dictated by the saturation vapor pressure above the liquid surface \citep{castan2011}.  The spatial variations in temperature---and hence vapor pressure---is the dominant source of atmospheric dynamics in its atmosphere.  
 
With observations of their atmospheres becoming available, it is therefore timely to produce model predictions for the magma oceans on lava planets.

\subsection{Interiors of Lava Planets}
The magma ocean on a lava planet is governed by the same geochemistry and geophysics as the magma oceans thought to have existed at the formation of rocky worlds in the Solar System \citep{2012AREPS..40..113E}, but with drastically different boundary conditions.  First of all, the magma ocean on a lava planet is in a steady state: it is not gradually cooling down.  Secondly, a lava planet's ocean is heated from above. Lastly, the magma ocean on a lava planet is horizontally heterogeneous, going from fully molten near the sub-stellar point ($T\approx 3000$) to completely solid near the day--night terminator ($T\approx 1000$)---in other words, lava planets cannot be treated as spherical cows.

Previous models of lava planet interiors assumed that the dayside magma ocean is shallow \citep[][]{leger2011,kite2016,nguyen2020}. Using a liquidus temperature gradient of about  30 K GPa$^{-1}$ \citep[][]{solomatov2000}, previous authors predicted that the temperature of the magma pond intersects the planet's liquidus temperature at a depth of about 70 km \citep[e.g., see appendix B of][]{kite2016}. 

However, for a multi-component system such as the bulk silicate Earth (BSE), the liquidus does not characterize the temperature at which magma becomes solid. The magmas composed of several oxides such as MgO, FeO, SiO$_2$ or CaO, fully solidify only when the temperature drops below the solidus. For temperatures between the solidus and the liquidus, the magma is partially molten. There is a rheological transition at about 50\% of melt fraction where magma starts to behave as a solid rather than a liquid  \citep[e.g.,][]{abe1993,lejeune1995,costa2005}. We show here that even though the temperature in the magma pond intersects the liquidus at shallow depth, the temperature in the deeper part of the planet remains close the liquidus due to latent heat effects.  

Major advances in laser-heated diamond anvil cell (LHDAC) now gives direct access to pressures and temperatures relevant to the deep interior of terrestrial planets. Recent measurements of liquidus and solidus temperature in LHDAC  \citep[e.g.,][]{fiquet2010,andrault2011,andrault2012,andrault2014,tateno2014,tateno2018,pradhan2015} show that the temperature difference between liquidus and solidus for Earth-like material can be as small as 400~K\citep{andrault2011} and as large 1000~K \citep{fiquet2010}. Using these recent experimental data at high pressures, we show that hemispherical magma oceans in lava planets are most likely deep and probably extend down to the core-mantle boundary for surface temperatures as low as 1900 K.

In section \ref{sec:model}, we present our assumptions and describe how we compute the interior temperature profile of a lava planet. In section \ref{sec:results} we present our results regarding the magma ocean depth, the location of the shore, and potential chemical gradients within the magma ocean due to partial crystallization. In section \ref{sec:discussion}, we speculate about the global internal dynamics of lava planets, its effect on atmospheric composition and future spectroscopic observations. 

\section{Model} \label{sec:model}

We consider a tidally locked planet of about 1.5 Earth radii that orbits close to its parent star. Since the overlying atmosphere is optically thin, the planet's surface temperature is well approximated by local radiative equilibrium \citep{castan2011,leger2009,nguyen2020}. The large angular size of the star as seen from the planet leads to an extended penumbra region and less than a hemisphere in the dark \citep{leger2011,nguyen2020} and we adopt a temperature at the sub-stellar point appropriate for K2-141b, $T_{\rm ss}=3000$~K. As we shall see, our model can easily apply to cooler lava planets. We also assume that the composition of the lava planet is that of the Bulk Silicate Earth \citep[BSE;][]{palme2003}.

On the day side, lava planets are expected to be fully molten \citep[e.g.,][]{kite2016} for surface temperature greater than 1700 K, forming a magma ocean. As temperature decreases toward the night side, the planetary surface is expected to transition from liquid to solid. The location of the shore is determined by (1) the variation of the surface temperature with respect to the angular distance from the sub-stellar point, and (2) the melting temperature of silicates at low pressure. Similarly, the depth of the magma ocean is constrained by (1) the interior temperature profile, and (2) the melting temperature of silicates at high pressures. 

The interior thermal structure of the planet has a first order control on the planet's dynamics as it governs the relative size of the magma ocean compared to the solid mantle. Both the depth and lateral extent of the magma ocean have a primary role in the atmosphere-interior coupling. Indeed, the rates of heat and mass transfer differ by several orders of magnitude between solid and liquid states. For the sake of simplicity, we assume that the planet's interior temperature follows an adiabat. This choice is motivated by the fact that the planet interior is expected to convect vigorosly. Convective motions are driven by temperature differences between the surface and the core-mantle boundary (CMB) as well as temperature contrast between the day-side and the night side. For comparison, the modern Earth has a surface temperature of about 300~K and CMB temperatures range between 4000 and 5000~K. For a planet with higher surface temperature such as lava planets, it is unlikely that the CMB temperature is lower than the Earth's at the same age. Hence the CMB temperature on a lava planet is expected to greatly exceed the surface temperature, even at the substellar point, certainly generating a flow of heat from the interior to the surface and possibly driving convective motions. In this study, we aim to estimate a pseudo steady-state interior thermal structure if heat is transported by convection. We neglect multiphase mechanical processes such as crystal settling, solidification dynamics, and melt percolation on the planet's structure \citep[e.g.,][]{maas2015,boukare2017a}.  
 
\subsection{Liquidus and solidus temperature} \label{subsec:liqsol}

Two temperatures are required to describe the melting behavior of silicate materials. Above the liquidus temperature, silicates are fully liquid. Below the solidus temperature, silicates are fully solid. Silicates are partially molten for temperatures that lie between the solidus and the liquidus. This is different from ``pure species'' such as SiO$_2$ or MgO whose phase change occurs at a single temperature. Both liquidus and solidus temperature are functions of pressure and composition. 

High pressure liquidus and solidus temperatures of the BSE are an area of active research  \citep{fiquet2010,andrault2011,andrault2012,andrault2014}. To propose a conservative estimate of the hemispherical magma ocean depth, we use the experimental data of \cite{fiquet2010}---the highest reported liquidus temperature. Our model thus provides a lower bound on the magma ocean depth. The experimental data of \cite{andrault2011}, for example, suggest a lower liquidus temperature and hence favor a deeper magma ocean.

\begin{figure*}[htb]
\plotone{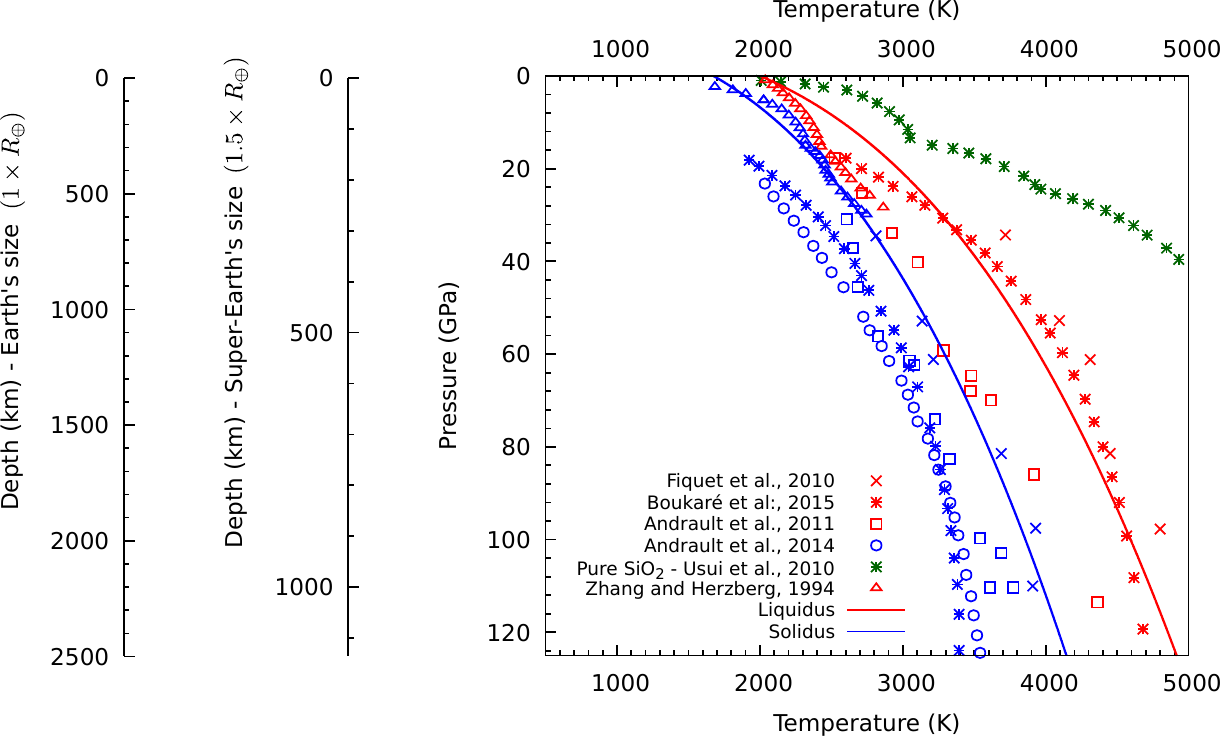}
\caption{High-pressure liquidus (red) and solidus (blue) temperature for Earth-like mantle compositions. Above the liquidus temperature, silicates are liquid. When the temperature lies between the liquidus and the solidus, silicates are partially molten. Silicates fully solidify when the temperature drops below the solidus.  Colored symbols show recent experimental measurements \citep{fiquet2010,andrault2011,andrault2014,zhang1994} and theoretical calculations \citep{boukare2015,usui2010} of the liquidus and solidus temperatures at high pressures. Discrepancies among data in the literature arise from differences in approaches such as silicate compositions or melt detection techniques. For the sake of simplicity, we approximate the liquidus and solidus curves for Earth-like composition by the red and blue lines, Equations (\ref{eq:tl}) and (\ref{eq:ts}). On the left, pressure-depth conversion is shown for the Earth and a $1.5 R_{\oplus}$ super-Earth. Note that the melting curve of pure SiO$_2$, shown by the  green asterisks, substantially overestimates the liquidus temperature of the Bulk Silicate Earth.}
\label{fig:data}
\end{figure*}

The solidus and liquidus temperatures are fitted to the experimental data of \cite{fiquet2010} and \cite{zhang1994}. The liquidus, $T_l$, and solidus, $T_s$, are plotted in Figure \ref{fig:data} and given by
\begin{equation}
	T_l(P) = 2000~{\rm K}\left(0.1169 \left(\frac{P}{\rm GPa}\right) +1\right)^{0.32726}
	\label{eq:tl}
\end{equation}
\begin{equation}
	T_s(P) = 1674~{\rm K}\left(0.0971 \left(\frac{P}{\rm GPa}\right) +1\right)^{0.351755}.
	%T_s(P) = 1674.0 \times (0.0971 \times  P +1)^{0.351755},
	\label{eq:ts}
\end{equation}
%where $P$ is the pressure in GPa.

\subsection{Isentropic temperature profiles} \label{subsec:isentrope}

The temperature profile in the fully molten region is calculated using the isentropic (and thus adiabatic) temperature gradient of a one phase system,
\begin{equation}
	\left(\frac{\partial T}{\partial P} \right)_{S}^{\rm op} = \frac{\alpha V}{C_p}T
	\label{eq:isentrope}
\end{equation}
where $\alpha$ is the thermal expansion, $V$ the volume, and $C_p$ the thermal capacity of the silicate melt. A self-consistent integration of Eq (\ref{eq:isentrope}) requires a thermal equation of state (EoS) for silicate melt. We use
\begin{eqnarray}
    	P(V,T)  = & 3 K_0f^{-2} (1-f)e^{\frac{3}{2}(K'-1)(1-f)}\nonumber\\ 
    	 & + \alpha_0 K_0 f^{3q-3}(T-T_0),
	\label{eq:pvt}
\end{eqnarray}
where the dimensionless length is $f= (V/V_0)^{1/3}$,
$K_0$ and $\alpha_0$ are the isothermal bulk modulus and thermal expansion at the reference conditions ($V_0$ and $T_0$), $K'=(\partial K_0/\partial P)_V$ and $q$ are constants. The first term on the right-hand side of Eq (\ref{eq:pvt}) corresponds to the Vinet-Rydeberg isothermal EoS \citep{vinet1987}, while the second term is the thermal pressure as initially proposed for solids \citep{anderson79}. Whereas the description of thermal pressure for solids requires positive values of $q$, liquids are generally described by negative values of $q$ \citep{asimow2010,boukare2015}. We fitted thermodynamic data of \cite{boukare2015} with Eq \ref{eq:pvt}. We use  $V_0=$ 26.27 cm$^3$ mol$^{-1}$, $T_0=$ 298  K, $\alpha_0 =  10.8 \times 10^{-5}$ K$^{-1}$, $K_0=$ 17.58 GPa, $K'=$ 6.9 and $q=$ -3.37.

In the partially molten region, the temperature profile follows a two-phase adiabat  \citep[e.g.,][]{miller1991,asimow1998} akin to moist adiabats in planetary atmospheres,
\begin{equation}
	\left(\frac{\partial T}{\partial P} \right)_{S}^{\rm tp} = \frac{\left(\frac{\partial T}{\partial P} \right)_{S}^{\rm hom} - \frac{T \Delta S}{C_p} \left(\frac{\partial F}{\partial T} \right)_P \left(\frac{\partial T_m}{\partial P} \right)_F }{1- \frac{T \Delta S}{C_p} \left(\frac{\partial F}{\partial T} \right)_P},
	\label{eq:bisentrope}
\end{equation}
where $\Delta S$ is the entropy difference between the liquid and the solid phase, $F$ is the melt fraction, $(\partial F/\partial T)_P$ is the variation of the melt fraction with temperature at constant pressure, and $(\partial T_m/\partial P)_F$ is the variation of the equilibrium temperature with pressure at constant melt fraction. Rigorous application of Eq (\ref{eq:bisentrope}) requires detailed self-consistent phase diagrams to constrain the latter parameters. Latent heat goes as $T \Delta S \propto T R $ and thermal capacity as $C_p \propto 3 R$, so we assume that $T\Delta S / C_p \propto T$, following \cite{miller1991}. To validate this assumption, we ran calculations using the thermodynamic model of \citet{boukare2015}, confirming that $\Delta S / C_p$ is of order unity. We use the crude assumption that melt fraction varies linearly between the liquidus and the solidus at fixed pressure, so that
\begin{equation}
    \left(\frac{\partial F}{\partial T} \right)_P = \frac{1}{T_l(P)-T_s(P)},
    \label{eq:linearf}
\end{equation}
where $T_l(P)$ and $T_s(P)$ are liquidus and solidus temperature at pressure $P$, respectively.

\section{Results} \label{sec:results}

\subsection{Magma ocean depth} \label{subsec:modepth}

\begin{figure*}[htb]
\plotone{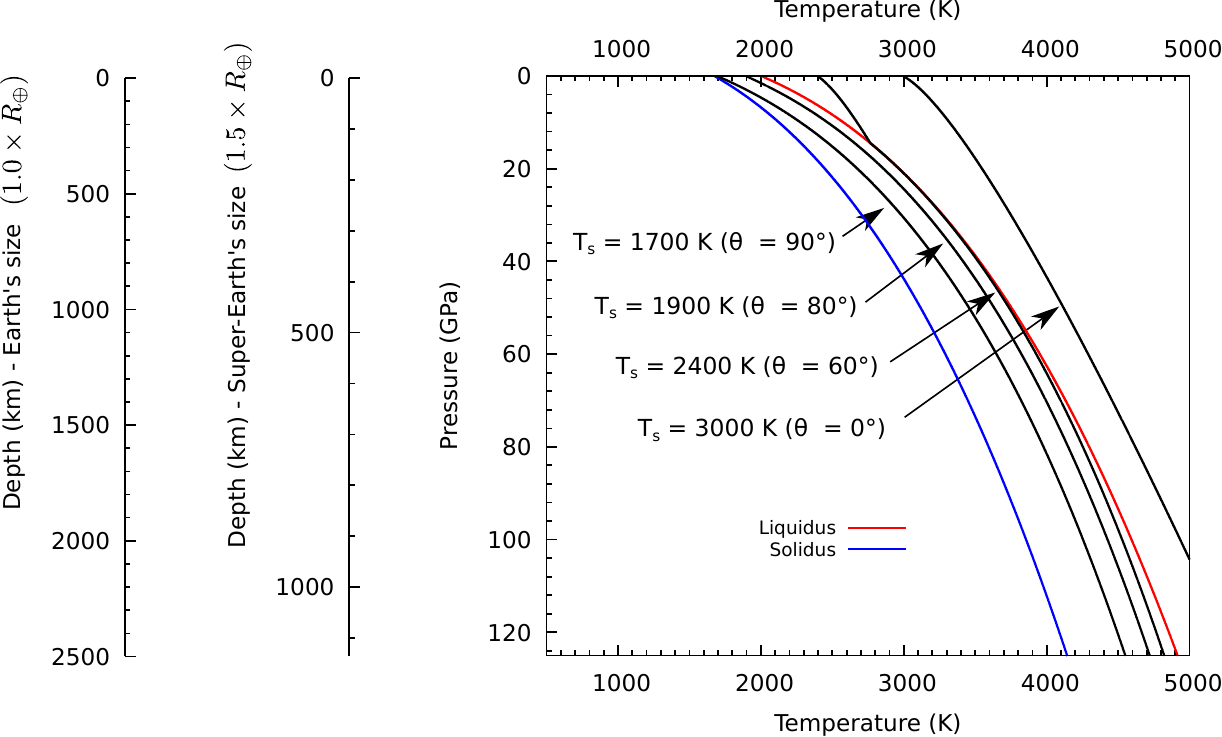}
\caption{Interior temperature of lava planets for four different surface temperatures: 1700 K, 1900 K, 2400 K, and 3000 K. The temperature increase with depth is due to the isentropic compression of the material. 
%The geotherms are computed as described in \cite{miller1991}, see eq. (\ref{eq:isentrope}) and (\ref{eq:bisentrope}). 
In regions hotter than the liquidus, the temperature follows a one-phase, homogeneous, isentrope (or adiabat). When the temperature is between the liquidus and the solidus, entropy conservation must account for the latent heat of fusion. In this region, the temperature follows a two-phase isentrope, akin to the moist adiabat of planetary atmospheres. For a surface temperature of 3000 K, the silicate mantle is expected to be fully molten for a planet the size of Earth or a $1.5 R_{\oplus}$ super Earth \citep[see also][]{solomatov2007tr,lebrun2013}. For a surface temperature of 2400 K, the geotherm intersects the liquidus at a depth of about 100 km for a $1.5 R_{\oplus}$ super Earth. However, the geotherms remains very close to the liquidus at greater depth such that the interior is still expected to be mostly molten. The lava planet interior behaves as a liquid, i.e. has a melt fraction greater than 50\%, for surface temperature above 1700 K (see Figure \ref{fig:meltfrac}). }
\label{fig:isentrope}
\end{figure*}

Our calculations show that the magma ocean on the permanent dayside of a lava planet is deeper than previously thought (see Figure~\ref{fig:isentrope}). For a surface temperature of 3000 K, the interior temperature remains above the liquidus temperature at least up 130 GPa, i.e, 1200~km in a $1.5 R_{\oplus}$ super-Earth. For a surface temperature of 2400~K, the interior temperature intersects the liquidus at about 15~GPa, i.e, 150~km in a $1.5 R_{\oplus}$ super-Earth, but it remains close to the liquidus at higher pressures. This is due to latent heat effects that bend adiabats such that they remain approximately parallel to the liquidus. The latter is controlled by the second term on the numerator of the right-hand side of  Eq~(\ref{eq:bisentrope}). For a surface temperature of 1700 K, the interior temperature is halfway in between the liquidus and the solidus temperature from the surface down to 120 GPa. These results are in excellent agreement with previous Earth's magma ocean model \citep[e.g.,][]{abe1997,solomatov2000,thomas2012,lebrun2013,monteux2016}.  

\begin{figure*}[htb]
\plotone{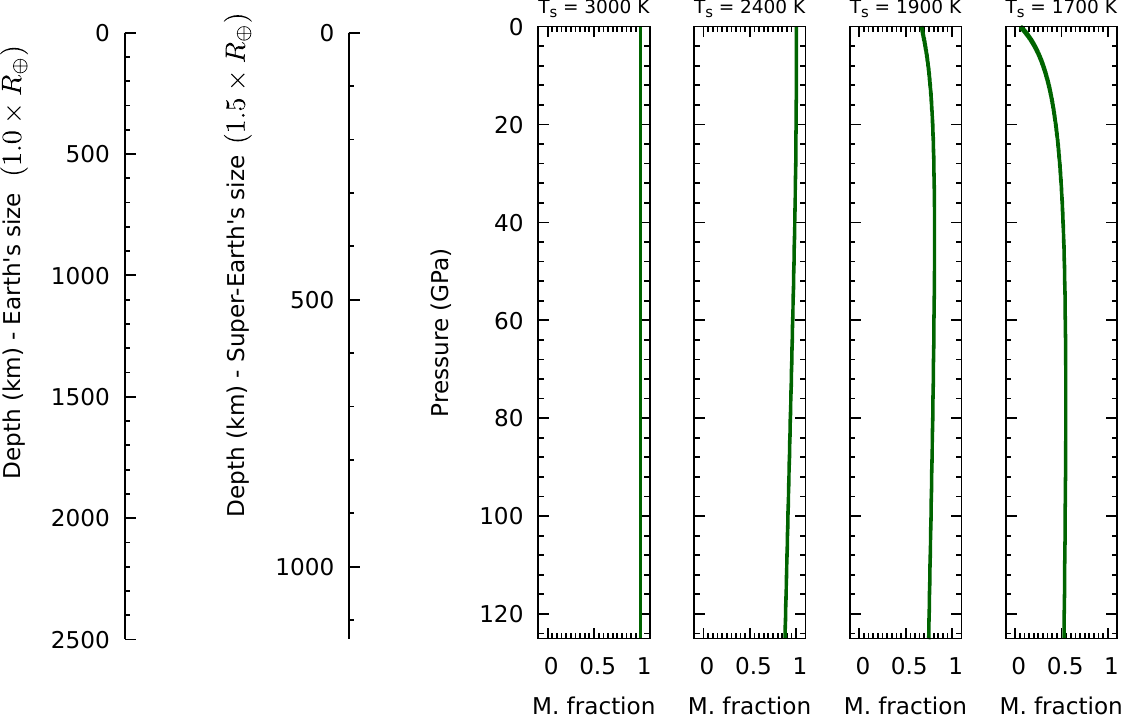}
\caption{Interior melt fraction as a function of depth for four surface temperatures: 3000 K, 2400 K, 1900 K, and 1700 K. Rheological transition between solid-like and liquid-like behavior occurs at about 50\% melt fraction.  
The bulk of the magma ocean therefore behaves as a sluggish fluid---not a solid---for surface temperature as low as 1700K. 
}
\label{fig:meltfrac}
\end{figure*}

We use Eq~(\ref{eq:linearf}) to determine the melt fraction as a function of depth for different surface temperature, shown in Figure~\ref{fig:meltfrac}.  Our calculations indicate that the melt fraction in the magma ocean stays above 50\% for surface temperatures hotter than 1700~K. As the rheologicical transition between solid and liquid occurs at about 50\% of melt fraction, these results show that the hemispherical magma ocean behaves as a liquid for surface temperature as low as 1700 K. For melt fraction close to the rheologicial transition, the multiphase mixture is expected to flow as a sluggish fluid---not as a solid. 

Our temperature and melt fraction profiles (Figures \ref{fig:isentrope} and \ref{fig:meltfrac}) can be used as a first-order estimation of the interior structure of lava planets in two cases. Our findings apply either to (1) multiple lava planets with different surface temperature at the sub-stellar point, or (2) to different locations in the same lava planet. If we neglect thermal mixing in the deep interior, these profiles can be used as a crude proxy for evaluating the effects on the interior thermal structure of lateral temperature variations imposed at the surface. Adopting this strong assumption for K2-141b, the interior profiles for surface temperatures of 1700 K, 2000 K, 2400 K, and 3000 K correspond to angular distances from the sub-stellar point of $\theta = 90$°, $\theta = 80$°, $\theta = 60$°, and $\theta = 0$°.

\subsection{Where is the magma ocean shore ? \label{subsec:shore}}

\begin{figure*}[htb]
\plotone{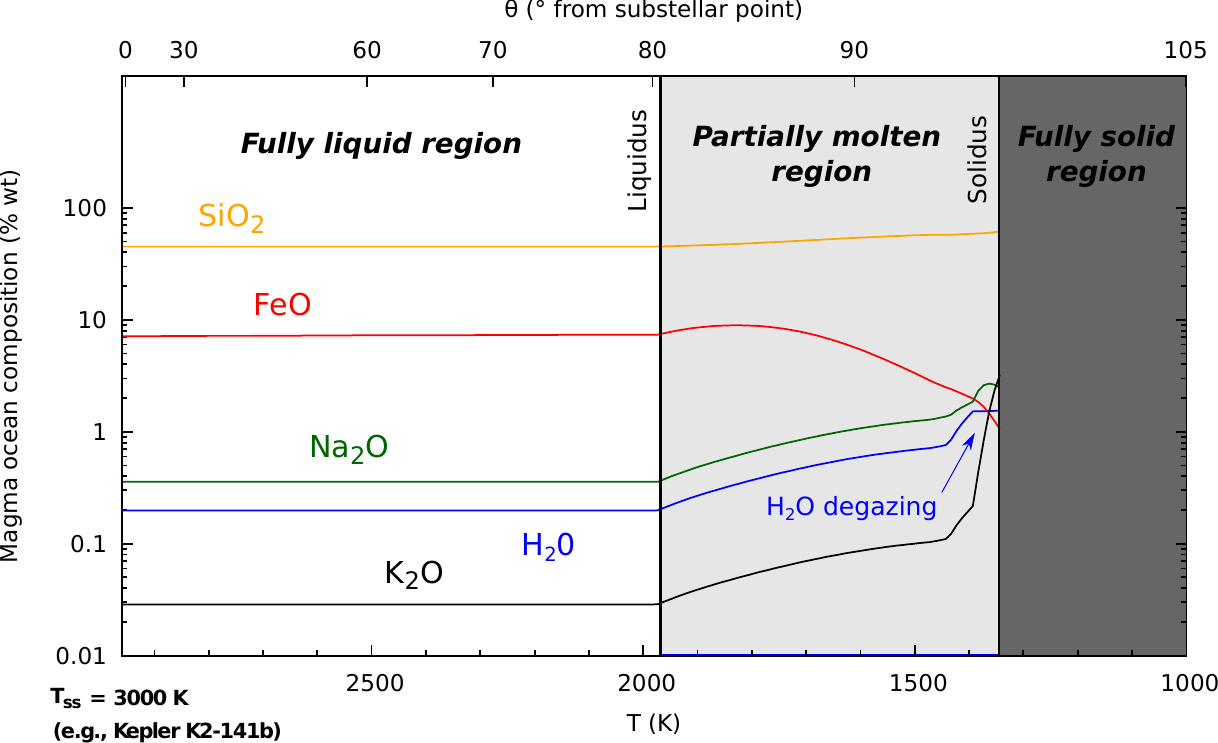}
\caption{Lateral variation of the surface composition of tidally-locked lava planets. Concentration of SiO$_2$ (yellow), FeO (red), Na$_2$O (green), H$_2$0 (blue), and K$_2$O (black) as a function of surface temperature (bottom axis) or distance from the sub-stellar point on K2-141b (top axis). There is no well-defined magma ocean shore but rather a gradual transition from the molten to the solid hemisphere. In this partially molten region, the composition of the magma evolves as precipitating minerals do not have the same composition as the melt.  Depending on the style of convective dynamics, these chemical heterogeneities can be mixed back into the convective region or permanently from the surface. For instance, enrichment in FeO would promote the formation of very dense melt that might sink to the bottom of the magma ocean \citep{kite2016}. Additionally, the concentration of volatile species such as H$_2$O and CO$_2$ can eventually reach the saturation limit in the melt near the solidus. When the melt is saturated, any increase in concentration results in the formation of bubbles that degas at the surface of the planet, with possible implications for transit observations of lava planets. 
}
\label{fig:chemvar}
\end{figure*}

Temperature decreases away from the sub-stellar point. Eventually, the temperature drops below the liquidus. We use the software MELTS 1.2.0 to compute the crystallization sequence of magma with a BSE composition at shallow depth, i.e., 250 bar. For the sake of simplicity, we only consider the case of batch crystallization where crystals remain in chemical equilibrium with the melt during solidification. These thermodynamic calculations allow us to predict more precisely the evolution of melt composition in major and minor elements as well as liquidus and solidus temperature at the surface of the planet. These temperatures constrain the location of the shore. 

For Kepler K2-141b, with a sub-stellar surface temperature of 3000~K, the transition from fully liquid to partially molten occurs near $\theta = 80^\circ$. The solidus temperature is reached at  $\theta = 100^\circ$. There is no well-defined shore at the surface of lava planets but rather a gradual transition from the molten to the solid hemisphere. 
It must be noted that the partially molten region will be larger for lava planets with more modest sub-stellar temperatures. For example, K2-22b has a sub-stellar surface temperature of 2000 K resulting in a partially molten region that extends from $\theta = 0^\circ$ to $\theta = 80^\circ$. 

\subsection{Chemical fractionation in the magma ocean \label{subsec:fractionation}}

Our calculations of the magma crystallization sequence (Figure \ref{fig:chemvar}) offers the possibility to predict compositional variation induced by solidification at the surface of a lava planet. Upon solidification, incompatible species stay in the melt instead of precipitating as minerals. Incompatible species such as Na$_2$O, K$_2$O or H$_2$O are thus concentrated in the melt as crystallization proceeds (see Figure  \ref{fig:chemvar}). Since melt fraction is primarily controlled by temperature, we propose that lateral variation of surface temperature generates compositional heterogeneities in lava planets by incongruent melt solidification. 

Moreover, our modeling indicates that an additional fluid phase composed mostly of H$_2$O exsolves from the melt as the temperature approaches the solidus. Here, H$_2$O will degas because its concentration in the melt has increased to the point of saturation. This can be seen in Figure~\ref{fig:chemvar} in the plateau of the magma water content close to the solidus temperature.  This process is different from atmosphere formation by liquid-vapor equilibrium where pressure is set by vapor exsolution itself. In the planet interior, pressure is set by the geostatic	 pressure. In that case, H$_2$O degassing is more akin to the behavior of salt in water. When water is saturated in salt, any increase in salt content result in salt exsolution.

\section{Discussion} \label{sec:discussion}

\subsection{Implications for lava planet interior structure and dynamics}

\begin{figure*}[htb]
\plotone{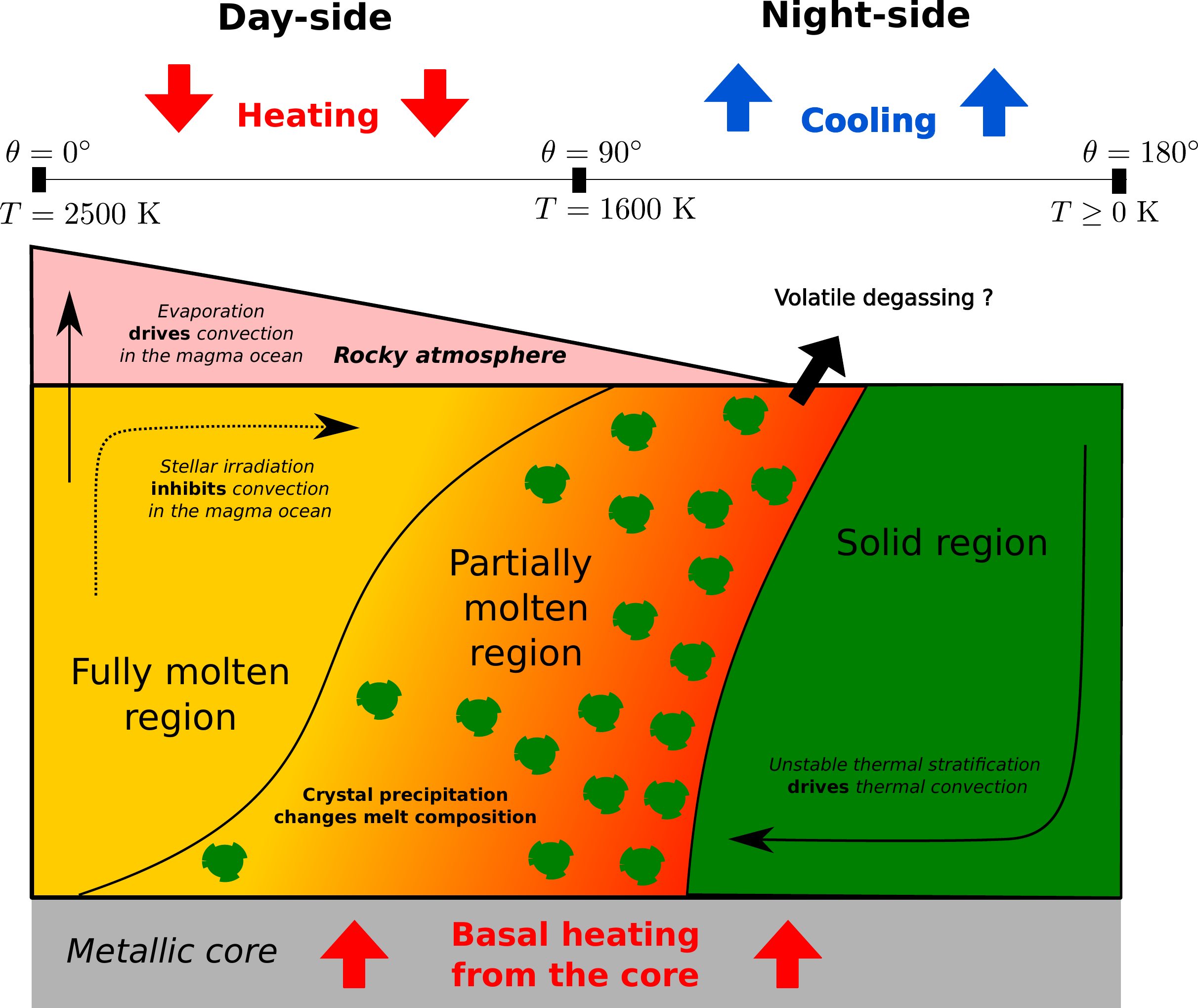}
\caption{Schematic of the interior structure and dynamics of a lava planet. The hemispherical magma ocean on the dayside could extend down to the core--mantle boundary. The night-side mantle is solid. A partially molten region probably connects the two hemispheres. The illustrated pseudo steady-state structure involves fluid motions. The solid-liquid interfaces are not static, rather rocky materials are expected to flow through the solid-liquid interface, melting or solidifying at these boundaries depending on the flow direction. The surface temperature gradient drives magma currents between the day and night sides. Temperature differences between core--mantle boundary and the planetary surface are expected to drive convection of the rocky mantle. Geodynamic models are required to explore the horizontal convective dynamics as well as the efficiency of solid-liquid phase separation and chemical mixing.}
\label{fig:cartoon}
\end{figure*}

Based on these thermodynamics calculations, we propose a new model for the interior structure of lava planets (figure \ref{fig:cartoon}). On the day side, the magma ocean is at least 1000~km deep, or 135~GPa, and probably extends down to the core-mantle boundary. Future high-pressure studies are required to constrain liquidus temperatures at conditions relevant for the deep mantles of Super-Earth, up to about 400~GPa for a $1.5 R_{\oplus}$ super-Earth with core-mantle ratio similar to the Earth. Crystal fraction increases progressively towards the night side, enriching the melt in incompatible elements. This process may be able to progressively differentiate the magma ocean depending on the efficiency of phase separation and chemical mixing in the interior. Compositional and density differences between melt and solids might have complex feedbacks with the interior fluid dynamics of lava planets. For instance, liquid silicates might become denser than their solids counterparts at high pressures \citep{funamori2010,boukare2015,boukare2017a,caracas2019}. Compared to previous studies that suggest a shallow magma ocean, our modeling points towards a deep hemispherical magma ocean that is expected to have profound implications for the atmosphere. Indeed, the composition of the atmosphere is dictated by the replenishment of surface material with fertile material coming from the magma ocean. 

Canonical magma ocean dynamics do not readily apply to tidally locked planets. The usual Rayleigh-Bénard convection in magma oceans, solid mantles and liquid cores develops in a fluid layer that is either cooled from above or heated from below \citep[e.g.,][]{solomatov2000}. In this regime, heat transport occurs vertically. For tidally locked lava planets, the temperature difference between the two hemispheres is expected to drive horizontal convection  \citep[e.g.,][]{hughes2008}. In this regime, heat transport proceeds horizontally similar to the thermohaline circulation in the Earth's oceans and the vertical temperature gradients is expected to differ from an adiabat. Indeed, thermal forcing at the surface results in thermal stratification in the bulk fluid, making the temperature profile subadiabatic in the dayside magma and superadiabtic on the night side. 

The high surface temperature of a lava planet does not necessary preclude the possibility of Rayleigh-Bénard-like convection. Indeed, the base of the rocky mantle is likely hotter the surface due to the slowly cooling iron core. Two scenarios could be envisioned in future studies depending on whether basal heating from the core is stronger compared to surface heating on the day-side. If basal heating is neglible compared to surface heating, the efficient cooling by advection on the night side may maintain a cold and solidified interior, on average. In that case, the present study probably overestimates the depth of the magma ocean as the interior temperature is not expected to follow an adiabat. We do not favor this hypothesis, however, as it appears unlikely that the temperature at the core-mantle boundary is lower than 3000~K. The surface temperature of the Earth is approximately 300~K and its CMB temperature is higher than 4000~K \citep[e.g.,][]{andrault2011,fiquet2010,labrosse2015}.

The maximum incoming heat flux on the day-side can be approximated by a conductive heat flux across a layer the size of the mantle. Using a thermal conductivity k = 1 W m$^{-1}$ K$^{-1}$, a temperature difference $\Delta T$ = 3000 K and mantle thickness H = 1.5 $R_{\oplus}$ /2 , we get a surface heat flux of about 0.3 TW. This surface heat flux is probably negligible compared to the secular cooling of the core that is of the order of 10 TW (today) to 60 TW (4 Gy ago) for the Earth  \citep[e.g.,][]{labrosse2015,nimmo2015}.
Consequently, basal heating of the mantle is expected to drive Rayleigh-Bénard convection and buffer the temperature profile to an adiabat, keeping the interior warm and molten on the day side.

Our calculations can be considered as a first order estimation that must be complemented by geodynamic models. However, this study raises interesting questions regarding lava planet’s interior dynamics. Fractionation by liquid-gas equilibrium at the surface has been proposed in several studies \citep[e.g.,][]{schaefer2009,kite2016}. Here, we emphasize the role of solid-liquid fractionation in the planet's interior. How do convection dynamics transition from the molten region to the solid hemisphere? How do compositional differences between melt and solid affect the interior dynamics ? How does magma ocean composition evolve with time? How do  mixing and fractionation compete to determine the steady-state compositon of the magma ocean? Since the atmosphere forms in equilibrium with the magma, any processes that control its composition at the surface are crucial for the atmosphere. Therefore, magma ocean interior dynamics must be considered as a fundamental aspect of lava planets atmospheres evolution. 

\subsection{Implications for (incompatible) volatile elements}

Models of magma oceans generally consider liquid--gas equilibrium at the planet surface as a primary mechanism for atmosphere formation  \citep[e.g.,][]{abe1986,bower2021}. However, vapor can also degas from the melt at depth when it becomes supersaturated in volatile, as observed in terrestrial volcanoes \citep[e.g.,][]{leguern1979erta,tazieff1994permanent}. This process might be an alternative to surface degassing when the magma ocean is isolated from the atmosphere by a solid conductive lid \citep[e.g.,][]{bower2021} and thus reinforces the efficiency of a lava planet to degas its volatile content in regions where the surface temperature is below the solidus, i.e., near the terminator and on the night-side. 

In the context of a rocky atmosphere, the fate of this putative volatile atmosphere is however uncertain. We expect strong winds from the day side to the night side. Studies of lava planet atmospheres have focused on Na because of its relative volatility \citep{castan2011,kite2016,nguyen2020}
or SiO because of its abundance in the 
mantle \citep{nguyen2020,zieba2022}. In either case, the atmosphere is expected to flow towards the day
—night terminator at velocities of kilometers per second, often exceeding the speed of sound. It has 
generally been assumed that any water present at the surface of the planet would be lost to space or 
cold trapped on the perpetually dark regions of the planet.  It is possible, however, that the gradual 
resurfacing of the nightside due to mantle overturn could resupply water to the magma ocean.  In that 
case, we predict that water would be degassed near the day—night terminator, only to be swept to the 
dayside of the planet by the background SiO wind. The presence of volatile species near the planet terminator could be tested in the near future with transit spectroscopy.

\section{Conclusion}
We have used state-of-the-art 1D models to make first-order predictions for the magma ocean on a lava planet. While these models are standard for Solar System studies, they have not yet been widely adopted in the context of exoplanets. The key to our results is the concept of partial melt and the high pressure thermodynamics of melting curves.

A rocky planet's mantle is a multi-component system and hence has a gradual transition from liquid to solid rather than the sharp phase transitions of pure substances such as water. As a result, a lava planet has no well-defined shore but rather a swath thousands of km wide where magma becomes increasingly sluggish.

The partial melting is also thermodynmically important: latent heat from the phase transition ensures that the isentrope is closer to isothermal and hence the transition to solid, or even sluggish liquid, occurs at higher pressures. We therefore predict that the magma ocean on K2-141b extends down to the core--mantle boundary.

Partial crystallization near the edge of the magma ocean leads to gradients in the surface composition of the melt and hence may affect the composition of the overlying rock vapor atmosphere. Near the mush--solid interface the melt becomes saturated in water, resulting in exsolution of water near the day--night terminator. Future work is needed to predict whether this could be a steady-state phenomenon or whether the water becomes permanently trapped as ice sheets on the planet's nightside.

Finally, we have adopted isentropic vertical temperature profiles in the magma ocean but the substantial horizontal gradients in temperature, chemistry,  evaporation and precipitation likely leads to magma ocean dynamics completely different from those considered in the Solar System context. Future challenges include numerically simulating the dynamics of lava planet magma oceans, and coupling such simulations to the atmospheric chemistry and dynamics that are probed by astronomical observations.

\section*{Acknoledgement}

We thank the anonymous reviewer for his/her insightful comments that
helped improve the paper. This work has received funding from the European Research Council (ERC) under the European Union's Horizon 2020 research and innovation program (grant agreement no. 101019965— SEPtiM). Parts of this work were supported by the UnivEarthS Labex program at Universit\'e de Paris and IPGP (ANR-10-LABX-0023 and ANR-11-IDEX-0005-02). We acknowledge the camaraderie and support of the McGill Space Institute and the "Institut de Recherche sur les exoplanètes" (IREX).

\bibliographystyle{aasjournal}



\end{document}